\documentclass[aps, prd, reprint, floatfix, showpacs,
preprintnumbers, superscriptaddress]{revtex4-1}

\usepackage{ac-tex}
\usepackage{xcolor}
\usepackage{hyperref}

\newcommand{\eq}[1]{Eq.~(\ref{eq:#1})}
\newcommand{\fig}[1]{Fig.~\ref{fig:#1}}
\newcommand{\tab}[1]{Table~\ref{tab:#1}}

\newcommand{\FF}[2]{{#1}_{#2}}
\newcommand{\Dirac}[1]{\FF{F_1}{#1}}
\newcommand{\Pauli}[1]{\FF{F_2}{#1}}
\newcommand{\SachsElec}[1]{\FF{G_E}{#1}}
\newcommand{\SachsMag}[1]{\FF{G_M}{#1}}

\newcommand{\FH}{Feynman--Hellmann}

\begin{document}

\title{Electromagnetic form factors at large momenta from lattice QCD}

\author{A.~J.~Chambers}
\affiliation{CSSM, Department of Physics,
 University of Adelaide, Adelaide SA 5005, Australia}
\email{alexander.chambers@adelaide.edu.au}

\author{J.~Dragos}
\affiliation{CSSM, Department of Physics,
 University of Adelaide, Adelaide SA 5005, Australia}
\affiliation{National Superconducting Cyclotron Laboratory
  and Department of Physics and Astronomy,
  Michigan State University, East Lansing, MI 48824, USA}

\author{R.~Horsley}
\affiliation{School of Physics and Astronomy,
 University of Edinburgh, Edinburgh EH9 3JZ, UK}

\author{Y.~Nakamura}
\affiliation{RIKEN Advanced Institute for Computational Science,
 Kobe, Hyogo 650-0047, Japan}

\author{H.~Perlt}
\affiliation{Institut f\"ur Theoretische Physik,
 Universit\"at Leipzig, 04103 Leipzig, Germany}

\author{D.~Pleiter}
\affiliation{JSC, J\"ulich Research Centre,
 52425 J\"ulich, Germany}
\affiliation{Institut f\"ur Theoretische Physik,
 Universit\"at Regensburg, 93040 Regensburg, Germany}

\author{P.~E.~L.~Rakow}
\affiliation{Theoretical Physics Division,
 Department of Mathematical Sciences,
 University of Liverpool, Liverpool L69 3BX, UK}

\author{G.~Schierholz}
\affiliation{Deutsches Elektronen-Synchrotron DESY,
 22603 Hamburg, Germany}

\author{A.~Schiller}
\affiliation{Institut f\"ur Theoretische Physik,
 Universit\"at Leipzig, 04103 Leipzig, Germany}

\author{K.~Somfleth}
\affiliation{CSSM, Department of Physics,
 University of Adelaide, Adelaide SA 5005, Australia}

\author{H.~St\"uben}
\affiliation{Regionales Rechenzentrum,
 Universit\"at Hamburg, 20146 Hamburg, Germany}

\author{R.~D.~Young}
\author{J.~M.~Zanotti}
\affiliation{CSSM, Department of Physics,
 University of Adelaide, Adelaide SA 5005, Australia}

\collaboration{QCDSF/UKQCD/CSSM Collaborations}
\noaffiliation{}

\pacs{12.38.Gc,14.20.Dh}

\preprint{
  ADP-17-06/T1012,
  DESY 17-019,
  Edinburgh 2017/02,
  LTH 1119
}

\begin{abstract}
  %
  Accessing hadronic form factors at large momentum transfers has
  traditionally presented a challenge for lattice QCD simulations.
  Here we demonstrate how a novel implementation of the
  \FH{} method can be employed to calculate hadronic form
  factors in lattice QCD at momenta much higher than previously
  accessible.
  Our simulations are performed on a single set of gauge
  configurations with three flavours of degenerate mass quarks
  corresponding to $m_\pi\approx 470 \text{ MeV}$.
  We are able to determine the electromagnetic form factors of the
  pion and nucleon up to approximately 6 GeV\textsuperscript{2}, with results for
  $G_E/G_M$ in the proton agreeing well with experimental results.
\end{abstract}

\maketitle

\section{Introduction}

One of the great challenges of hadron physics is to build
consistent and informative pictures of the internal structures
of strongly-interacting particles.
An important aspect of this endeavour is the
calculation of electromagnetic form factors for various baryons and mesons.
These encode a description of
the distribution of electromagnetic currents in hadrons and are key to
describing the extended structure of these composite states.

For most of the second half of the 20\textsuperscript{th} century, measurements of the
electromagnetic form factors of the nucleon were
obtained using the Rosenbluth separation technique~\cite{Rosenbluth:1950yq} (also e.g.~\cite{Qattan:2004ht}).
Broadly, these experiments indicated that the electric and magnetic form factors
scaled proportionally for $Q^2$ up to around 6 GeV\textsuperscript{2}, with
$\mu_p \SachsElec{p}/\SachsMag{p} \approx 1$.
This was later found to be in
disagreement with recoil polarisation experiments at Jefferson Lab which showed $\mu_p
\SachsElec{p}/\SachsMag{p}$ decreasing approximately linearly for $Q^2 \gtrsim 0.5
\text{ GeV}^2$ (see
e.g.~\cite{Jones:1999rz,Gayou:2001qd,Punjabi:2005wq,Puckett:2010ac, Puckett:2011xg}).
This discrepancy is now largely understood through studies of two-photon exchange
effects in the Rosenbluth method~\cite{Guichon:2003qm,Blunden:2005ew}.
%
Nevertheless, it is still unknown whether the linear $Q^2$ trend
continues and crosses zero, or if the fall-off with $Q^2$ slows down.
Experimental results are as yet unable to obtain precise results at
the relevant momentum scales, and so this remains an open question.
Resolving the scaling of the form factors in this domain is one of the key
physics goals of the upgraded CEBAF at Jefferson Lab.

The large-$Q^2$ behaviour of the pion electromagnetic form factor
$F_\pi$ has proven challenging to probe experimentally --- see
Refs.~\cite{Volmer:2000ek,Horn:2006tm,Huber:2008id} for recent
innovative advances.
Besides providing information about the electromagnetic structure of
the pion, the $Q^2$-behaviour of $F_\pi$ provides insight into the
transition from the soft to the hard regime in QCD (see
\cite{Chang:2013nia} for a recent example).
Owing to the present limitations, experimental data are unable to reliably
discriminate different models describing the transition to the
asymptotic domain~\cite{Horn:2016rip}.

Lattice calculations of hadronic form factors have typically focussed on
the study of processes at low-momentum transfer (see
e.g.~\cite{Collins:2011mk,Alexandrou:2011db,Shanahan:2014uka,Shanahan:2014cga,Green:2014xba,Capitani:2015sba}),
with only limited studies at large $Q^2\gtrsim 3 \text{ GeV}^2$~\cite{Lin:2010fv,Koponen:2017fvm}.
%
%
There are a variety of reasons that have contributed to the diffculty
in accessing high-momentum transfer in lattice QCD. Given that the form
factors fall with $Q^2$, it is immediately clear that one is attempting
to extract a much weaker signal from datasets obtained with finite
statistics.
Further, in terms of the numerical computation, the signal-to-noise
ratio of hadron correlators rapidly deteriorates as the momentum of
the state is increased.
This had commonly led to the study of 3-point correlators which are
projected to zero momentum at the hadron sink.
In this case, the possible momentum transfers are limited by the
maximum momentum available at the source.
With limited statistical signal, it is therefore difficult to assess
the degree of excited-state contamination, which can lead to
significant systematic uncertainty
\cite{Lin:2010fv,Owen:2012ts,Green:2014xba,Yoon:2016dij,Dragos:2016rtx}.

In the present work we demonstrate the ability to access high-momentum
transfer in hadron form factors on the lattice using an extension of
the \FH{} theorem to non-forward matrix elements.
This builds upon recent applications of the \FH{} theorem
for hadronic matrix elements in lattice QCD~\cite{Horsley:2012pz,Chambers:2014qaa,Chambers:2014pea,Chambers:2015bka}
--- see also
Refs.~\cite{Detmold:2004kw,Engelhardt:2007ub,Detmold:2009dx,
  Detmold:2010ts,Primer:2013pva,Freeman:2014kka,Savage:2016kon,Bouchard:2016heu}
for similar related techniques.
Through the \FH{} theorem one relates matrix elements to
energy shifts.
In the case of lattice QCD this allows one to access matrix elements
from 2-point correlators, rather than a more complicated analysis of
3-point functions.
This greatly simplifies the process of neutralising excited-state
contamination.
As described below, the method most naturally works in the Breit frame
($E(\vec{p}')=E(\vec{p})$) and hence one maximises the
momentum transfer for any given accessible state momentum $|{\bf p}|$.
Finally, the high degree of correlations in the gauge ensembles makes
it possible to extract a weak signal from a relatively noisy state.


\section{\label{sec:fh_methods}\FH{} Methods}
To extend the \FH{} analysis to non-forward matrix elements,
we first consider a simple quantum mechanical situation.
The familiar form of the FH{} theorem reads
\begin{equation}
\dpd{E_\psi}{\lambda}=\braopket{\psi}{\dpd{H}{\lambda}}{\psi}
\eqcomma
\end{equation}
where $E$ is the energy eigenvalue of the state $\psi$.
This readily follows from first-order perturbation theory.
In the presence of spatially-varying external fields, the
conventional theorem requires a slight modification.
We consider some first-order perturbation of the Hamiltonian, $H=H_0+\lambda V$, which couples
to a definite (real) spatial Fourier component
\begin{equation}
\frac{\partial H}{\partial \lambda}\equiv
\widetilde{V}_+(\vec{q})=\widetilde{V}(\vec{q})+\widetilde{V}(-\vec{q})
\eqcomma
\end{equation}
defined in terms of the complex Fourier modes
$\widetilde{V}(\vec{q})=\int \mathrm{d}^3 \vec{y}\,
e^{i\vec{q}.\vec{y}}V(\vec{y})$,
for some Hermitian potential $V(\vec{y})$.
The diagonal matrix elements of this operator vanish in the basis of
definite momentum eigenstates
\begin{equation}
\langle \vec{p}|\widetilde{V}_+(\vec{q})|\vec{p}\rangle=0
\eqcomma
\end{equation}
and standard perturbation theory would suggest that there is no
shift of the energy level at first order in $\lambda$. The exception
to this rule is in the case of a degeneracy in the unperturbed
eigenstates $E_0(\vec{p})=E_0(\vec{p}\pm\vec{q})$.
The familiar solution in this case is to invoke degenerate perturbation theory
where one diagonalises the space of the degeneracy with respect
to the applied external potential. The degeneracy condition dictates
that one is considering Breit-frame transitions. For demonstrative
purposes, we choose the simple case where $\vec{p}=\pm\vec{q}/2$ and hence
at lowest order in the field strength
the system is diagonalised by the states
$|\vec{q}/2\rangle_\pm\propto |\vec{q}/2\rangle\pm|-\vec{q}/2\rangle$.
The corresponding eigenvalues are given by $E_0(\vec{q/2})\pm\lambda \Delta E+\mathcal{O}(\lambda^2)$, where the energy shift corresponds to the matrix element of interest,
\begin{align}
  \Delta E &= {}_+\langle
             \vec{q}/2|\widetilde{V}_+(\vec{q})|\vec{q}/2\rangle_+
  =\langle \vec{q}/2|\widetilde{V}(\vec{q})|-\vec{q}/2\rangle
             \eqstop
\end{align}

Owing to the discretised spectrum (and momentum) on the lattice,
this quantum mechanical argument translates in a straightforward
fashion to hadronic matrix elements. In the case of continuous
momenta the presence of the periodic potential induces a gap in
the dispersion curve, as in conventional band theory.

%
To implement within a lattice calculation, the Lagrangian is
modified to incorporate a spatially-varying external potential
\begin{align}
  \Lagrangian(y)
  & \to
  \Lagrangian(y)
  + \lambda \left(
    e^{+i \vec{q} \cdot \vec{y}}
    + e^{-i \vec{q} \cdot \vec{y}}
    \right)
  \mathcal{O}(y)
    \eqcomma
  \label{eq:general_lag_mod}
\end{align}
where the phase of the exponentials is defined with respect to the
location of the hadron source at $\vec{y}=\vec{0}$.
The symbol $\mathcal{O}$ denotes a quark-bilinear operator and $\lambda$
represents the strength of the external field --- which is kept small
ensure that the energy response is in the linear regime.
Alternatively, one may isolate the linear $\lambda$ dependence of the
correlator directly by constructing compound propagators
\cite{Savage:2016kon,Bouchard:2016heu}.

To compute connected quark contributions, quark propagators are inverted
according the modified action corresponding to \eq{general_lag_mod} --- sea-quark
contributions would require new gauge ensembles~\cite{Chambers:2015bka}
or an effective reweighting technique.
Fourier-projected, hadron correlation functions are defined by
\begin{align}
C_{\vec{p}}^\lambda(t)=\sum_{\vec{x}}e^{-i\vec{p}.\vec{x}}\,
{}_\lambda\langle \Omega | \chi(t,\vec{x})\chi^\dagger(0,\vec{0})
|\Omega\rangle_\lambda
  \eqcomma
\end{align}
where subscript $\ket{\Omega}_\lambda$ is the vacuum of the modified theory.
The spectrum can be directly isolated by
constructing even and odd linear combinations,
\begin{align}
C_{\vec{p},\vec{p}'}^{\lambda\pm}=C_{\vec{p}}^\lambda\pm
  C_{\vec{p}'}^\lambda
  \eqcomma
\end{align}
of Breit-frame momentum pairs, $\vec{p}$ and $\vec{p}'(=\vec{p}+\vec{q})$.
To isolate an energy shift, it is more straightforward to implement the
``$+$'' combination $C_{\vec{p},\vec{p}'}^{\lambda+}$ rather than the ``$-$''
sum, which vanishes in the free-field limit.

Only the Breit-frame pairs will receive an energy shift which is linear in
the applied field strength $\lambda$. This energy shift corresponds directly
to the hadronic matrix element of interest
\begin{equation}
  \evalat{\dpd{E_H(\vec{p}')}{\lambda}}_{\lambda=0}
  =
  \frac
  {\braopket{H(\vec{p}')}
    {\mathcal{O}(0)}
    {H(\vec{p})}}
  {\braket{H(\vec{p}')}{H(\vec{p}')}}
  \label{eq:fh_spinzero}
  \eqcomma
\end{equation}
or similarly for $\vec{p} \leftrightarrow \vec{p}'$.
We have confirmed numerically that non-Breit frame states do not
receive a linear energy response, as expected.



\section{\label{sec:sim_deets}Simulation Details}

In the present work, we use an ensemble of 1700 gauge field configurations
with $2+1$ flavours of non-perturbatively $O(a)$-improved Wilson fermions and a
lattice volume of $L^3 \times T = 32^3 \times 64$.
The lattice spacing $a = 0.074(2) \text{ fm}$ is set using a number of singlet
quantities~\cite{Horsley:2013wqa, Bornyakov:2015eaa,
  Bietenholz:2010jr, Bietenholz:2011qq}.
The clover action used comprises the tree-level Symanzik improved
gluon action together with a stout smeared fermion action, modified
for the implementation of the FH{} method~\cite{Chambers:2014qaa}.

We use a single ensemble with hopping parameters,
$(\kappa_l, \kappa_s) = (0.120900, 0.120900)$, which correspond to a
pion mass of $\sim 470 \text{ MeV}$.
To study electromagnetic form factors, quark propagators are calculated with the modified Lagrangian
\begin{align}
    \Lagrangian(y) \to
  \Lagrangian(y) +
  \left(
    e^{+i \vec{q} \cdot \vec{y}}
    + e^{-i \vec{q} \cdot \vec{y}}
    \right)
  \bar{q}(y) \lambda \cdot \gamma q(y)
  \eqcomma
  \label{eq:vector_lag_mod}
\end{align}
for multiple values of $\vec{q}$ as listed in \tab{momenta},
where either $\lambda_2$ or $\lambda_4$ take non-zero values of $1
\times 10^{-4}$ or $-1 \times 10^{-5}$.
Note that we only use the simplest Breit-frame kinematics, $\vec{p}' =
- \vec{p}$.
%
This choice allows us to minimise
${\vec{p}}^2$ for each value of $\vec{q}^2$, and hence
minimise the noise in the correlator. As described below, this also
projects the nucleon energy shifts directly onto $G_E$ or $G_M$.
\begin{table}
  \centering
  \begin{tabular}{c c c c}
    \hline
    \hline
    $\vec{q}$ & $\vec{p}$ & $\vec{p}^2$ & $Q^2$ \\
    \hline
    $(0,0,0)$ & $(0,0,0)$ & $0$ & $0$   \\
    $(2,0,0)$ & $(1,0,0)$ & $1$ & $4$   \\
    $(2,2,0)$ & $(1,1,0)$ & $2$ & $8$   \\
    $(2,2,2)$ & $(1,1,1)$ & $3$ & $12$  \\
    $(4,0,0)$ & $(2,0,0)$ & $4$ & $16$  \\
    $(4,2,0)$ & $(2,1,0)$ & $5$ & $20$  \\
    $(4,2,2)$ & $(2,1,1)$ & $6$ & $24$  \\
    \hline
    \hline
  \end{tabular}
  \caption{Momentum insertions and the corresponding Breit-frame momenta
    used in these calculations, where $\vec{p}'=-\vec{p}$.  Momenta
    are given in lattice Fourier units of $2\pi/L$.}
  \label{tab:momenta}
\end{table}

\section{\label{sec:results}Results}


\subsection{Electromagnetic Form Factors of the Nucleon}
The (Euclidean) decompositon of the vector current for the individual
quark flavour contributions of the nucleon is written in terms of the
familiar Dirac and Pauli ($\Dirac{q}$ and $\Pauli{q}$) form factors,
\begin{align}
  & \braopket
  {N(p', s')}
  {\bar{q}(0) \gamma_\mu q(0)}
  {N(p, s)}
    =
    \notag \\
  &
  \bar{u}(p', s')
  \left[
  \gamma_\mu \Dirac{q} (Q^2)
  + \frac{\sigma_{\mu \nu} q_\nu}{2 M_N} \Pauli{q} (Q^2)
  \right]
  u(p, s)
  \eqcomma
\end{align}
where we denote the invariant 4-momentum transfer squared as $Q^2=-q^2=-{(p'-p)}^2$.
The Sachs electromagnetic form factors are defined by
\begin{align}
  \SachsElec{q} & = \Dirac{q} - \frac{Q^2}{{(2M)}^2} \Pauli{q}
                  \\
  \SachsMag{q} & = \Dirac{q} + \Pauli{q}
\end{align}

For the incident-normal Breit frame ($\vec{p}'=-\vec{p}$), the temporal
and spatial components of the current give rise to energy shifts which
directly project out the electric and magnetic form factors respectively,
\begin{align}
  \evalat{\dpd{E_N}{\lambda_4}}_{\lambda=0}
  & \overset{\vec{p}' = -\vec{p}}{=} \frac{M_N}{E_N} \SachsElec{q}
  \eqcomma
  \\
  \evalat{\dpd{E_N}{\lambda_i}}_{\lambda=0}
  & \overset{\vec{p}' = -\vec{p}}{=}
  \frac{{[\uvec{e} \times \vec{q}]}_i}{2E_N} \SachsMag{q}
  \eqcomma
\end{align}
where $\uvec{e}$ is the spin polarisation vector determined by the
choice of polarisation direction of the nucleon.

\begin{figure}
  \includegraphics[width=\columnwidth]{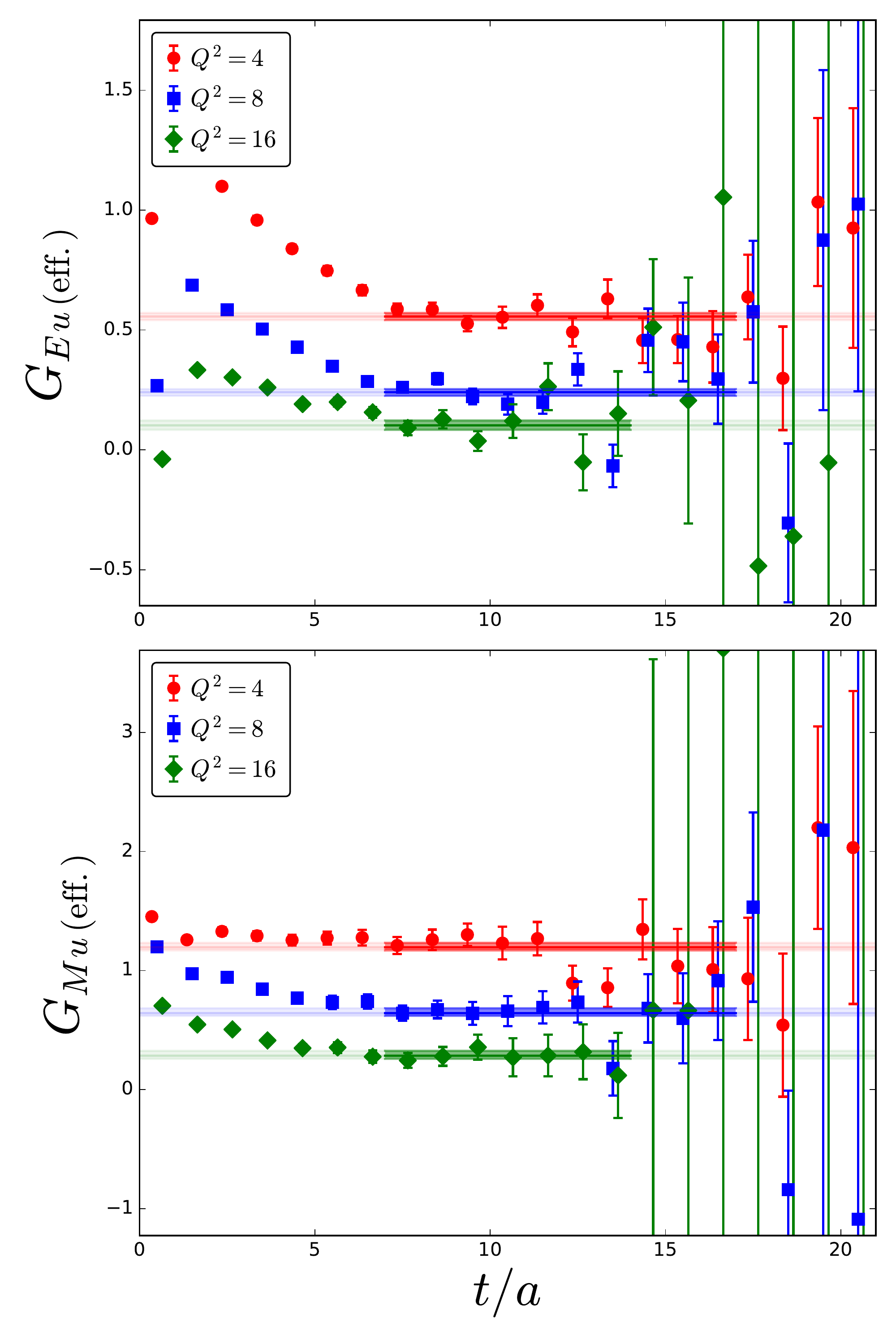}
  \caption{Effective electric and magnetic form factors of the $u$
    quark in the nucleon for different values of $Q^2$. Results shown
    are for a single value of $\lambda \ne 0$ (since we are in the
    linear region, results at
    different $\lambda$ are statistically identical).}
\label{fig:effective_G_E_u}
\end{figure}

Utilising ratios of correlators with and without the applied
external field, we can defined ``effective form factors'' by
appropriate scaling of the effective energy shift $\Delta{E_N}_{\text{eff.}}$,
\newcommand{\EffSachsElec}[1]{{\SachsElec{#1}}_{(\text{eff.})}}
\newcommand{\EffSachsMag}[1]{{\SachsMag{#1}}_{(\text{eff.})}}
\begin{align}
  \EffSachsElec{q}
  & = \frac{E_N}{M_N}
    \frac{\Delta {E_N}_{\text{(eff.)}}}{\lambda}
    \eqcomma \\
  \EffSachsMag{q}
  & = \frac{2 E_N}{{[\uvec{e} \times \vec{q}]}_i}
    \frac{\Delta {E_N}_{\text{(eff.)}}}{\lambda_i}
    \eqstop
\end{align}
These should plateau to the relevant form factors provided $\lambda$
is small enough that the energy shift is predominantly linear.
\fig{effective_G_E_u} shows results for effective electromagnetic form factors
for a subset of $Q^2$ values.
Here we identify that quite clean plateaux are realised up to quite
large momentum transfer. As a check on the selected fit window,
we ensure that the free-field correlators are sufficiently
saturating to the ground-state energy dispersion.

\begin{figure}
  \includegraphics[width=\columnwidth]{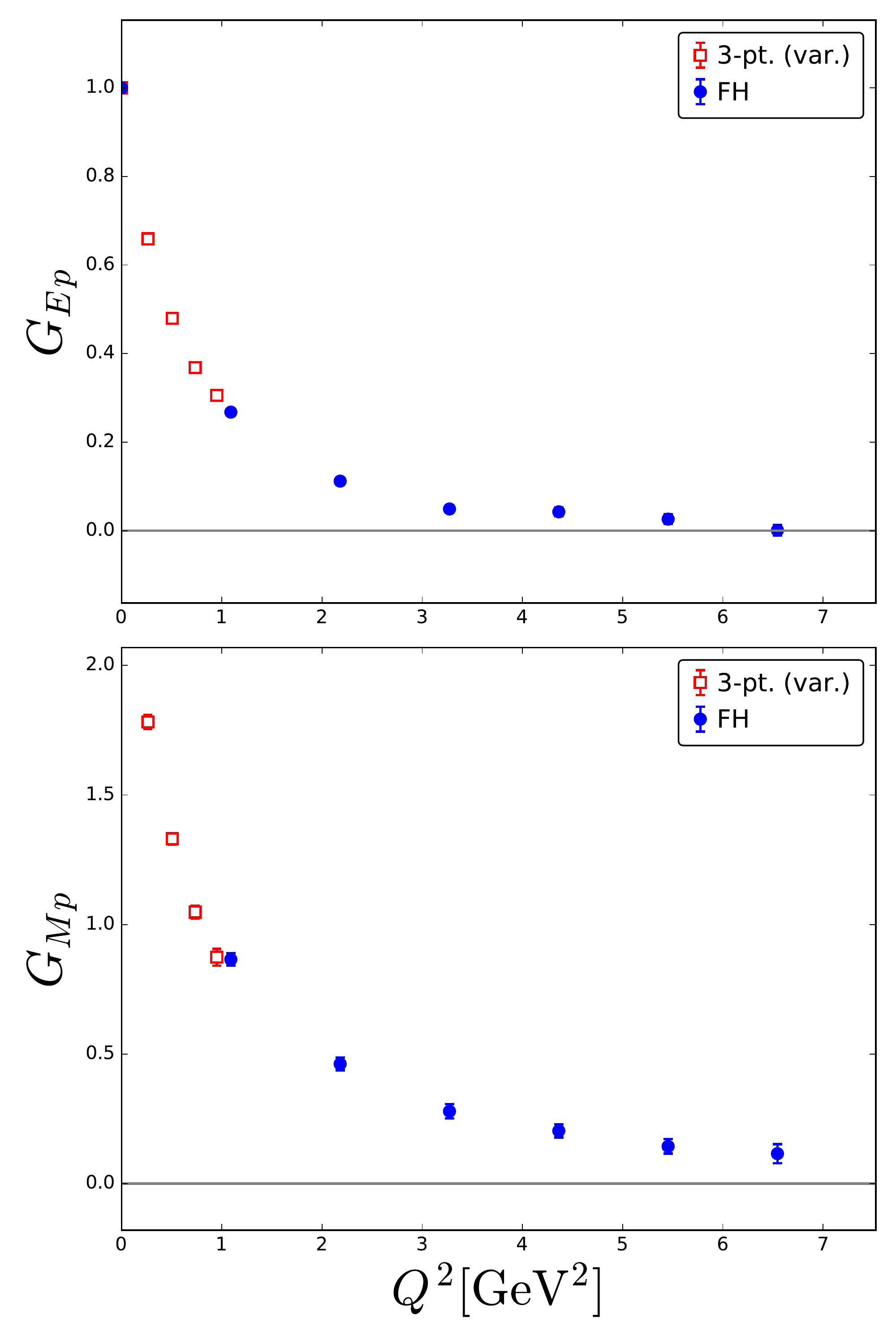}
  \caption{$G_E$ and $G_M$ for the proton from the \FH{}
    method and a variational method described in~\cite{Dragos:2016rtx}
  employed on the same ensemble.}
\label{fig:G_EM_p}
\end{figure}

\fig{G_EM_p} shows results for the proton electric
and magnetic form factors ---
neglecting disconnected contributions,
which are anticipated to be very small at large $Q^2$~\cite{Green:2015wqa}.
In the low-$Q^2$ region we compare with results computed on the same
ensembles using a variationally-improved 3-point function approach,
as described in~\cite{Dragos:2016rtx}. Very good agreement is
observed in the region of overlap. The statistical signal for the
new \FH{} approach is seen to extend to much larger
$Q^2$ than has been accessible in the past.

Phenomenologically, the $Q^2$-range we are now able to access would
allow for tighter constraints to be placed on the distribution of
charge and magnetisation in the nucleon at small impact parameter
\cite{Venkat:2010by}.

\begin{figure}
  \includegraphics[width=\columnwidth]{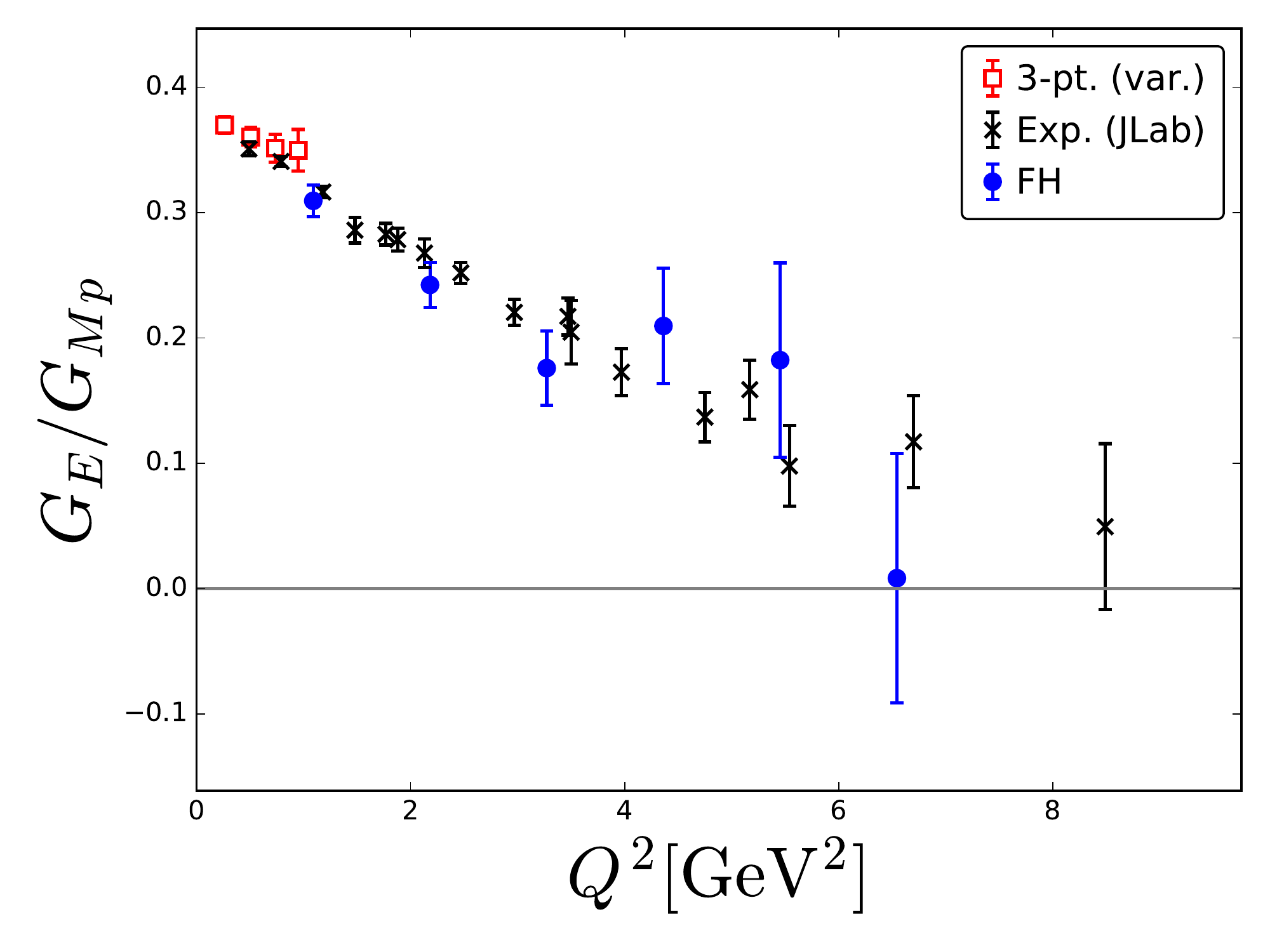}
  \caption{Ratio $G_E/G_M$ for the proton from application of the
    \FH{} method, from a variational analysis of
    three-point functions~\cite{Dragos:2016rtx}, and from experiment~\cite{Punjabi:2005wq, Puckett:2010ac, Puckett:2011xg}. Note this is not scaled by
    the magnetic moment of the proton $\mu_p$, as this would require
    phenomenological fits to the low $Q^2$ data, which is not the
    focus of this work.}
\label{fig:ratioGEGM}
\end{figure}

\fig{ratioGEGM} displays the extraction of
the ratio $G_E/G_M$ as a function of $Q^2$, and a comparison to experiment~\cite{Punjabi:2005wq, Puckett:2010ac, Puckett:2011xg}.
While nothing definitive can be concluded about a potential zero
crossing, the overall trend is seen to compare very well with the
experimental data.

\subsection{Electromagnetic Form Factor of the Pion}

\begin{figure}
  \includegraphics[width=\columnwidth]{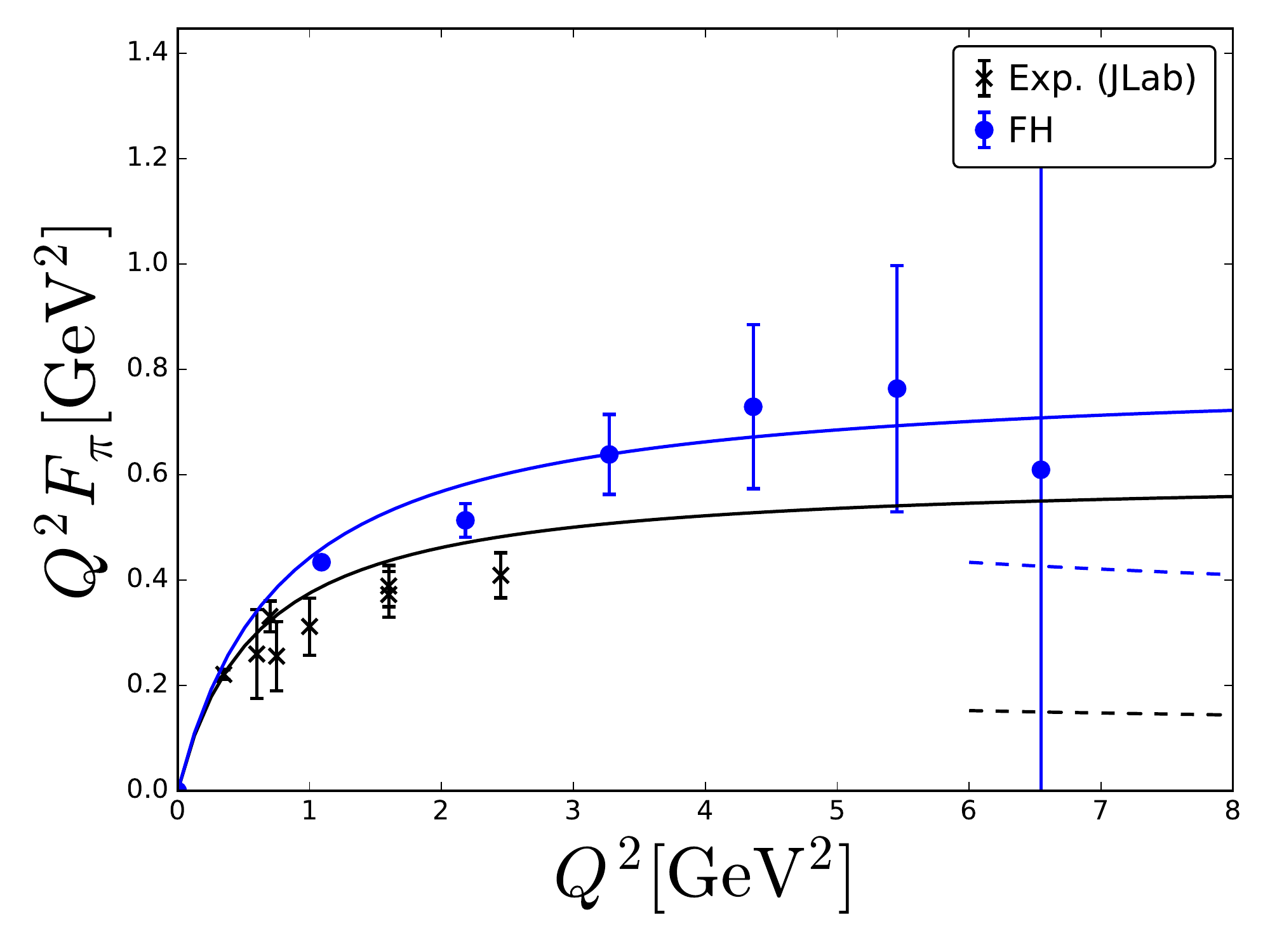}
  \caption{Scaled pion form factor $Q^2F_\pi$ from the
   \FH{} technique and from
    experiment~\cite{Huber:2008id}. The solid lines are the vector meson
    dominance at the relevant pion masses, and the dotted lines are
    the asymptotic values predicted by perturbative QCD
    (see~\cite{Chang:2013nia} for a discussion of this value and its
    limitations).}
\label{fig:pion_exp_comp}
\end{figure}

Following a similar analysis as that for the nucleon, we show the
determination of the pion form factor and comparison to experiment~\cite{Huber:2008id} in \fig{pion_exp_comp}.
The realised statistical signal gives confidence that future lattice
simulations will be able to provide important insight into this transition
between the perturbative and nonperturbative.

\section{\label{sec:conclusion}Conclusion}

In this work we have extended the \FH{}
technique to access non-forward matrix elements.
We demonstrate that this provides for a dramatic improvement in the
ability to extract nucleon and pion form factors at much higher
momentum transfers than previously possible.
Before making rigorous comparisons with phenomenology, standard
lattice systematics must be further quantified, including quark mass
dependence, discretisation artifacts and continuum extrapolation.
There is also further potential for increased precision by using
improved operators that have better access to high-momentum states,
as proposed in~\cite{Bali:2016lva}.

The high-momentum form factors extracted in this work demonstrate
a significantly expanded scope for lattice QCD to address this
phenomenologically interesting domain of hadron structure.

\acknowledgments{}

The numerical configuration generation was performed using the BQCD
lattice QCD program~\cite{Nakamura:2010qh}, on the IBM BlueGeneQ
using DIRAC 2 resources (EPCC, Edinburgh, UK), the BlueGene P and Q at
NIC (J\"ulich, Germany) and the Cray XC30 at HLRN
(Berlin--Hannover, Germany).
Some of the simulations were undertaken using resources awarded at the
NCI National Facility in Canberra, Australia, and the iVEC facilities
at the Pawsey Supercomputing Centre. These resources are provided
through the National Computational Merit Allocation Scheme and the
University of Adelaide Partner Share supported by the Australian
Government.
This work was supported in part through supercomputing resources
provided by the Phoenix HPC service at the University of Adelaide.
The BlueGene codes were optimised using Bagel~\cite{Boyle:2009vp}.
The Chroma software library~\cite{Edwards:2004sx}, was used in the
data analysis.
AC was supported by the Australian Government Research Training Program Scholarship.
JD gratefully acknowledges support by the National Superconducting Cyclotron Laboratory (NSCL)/Facility
GS was supported by DFG grant SCHI 179/8-1.
HP was supported by DFG grant SCHI 422/10-1.
for Rare Isotope Beams (FRIB) and Michigan State University (MSU) during the preparation of this work.
This investigation has been supported by the Australian Research
Council under grants FT120100821, FT100100005 and DP140103067
(RDY and JMZ).

\bibliography{ref}

\begin{thebibliography}{47}%
\makeatletter
\providecommand \@ifxundefined [1]{%
 \@ifx{#1\undefined}
}%
\providecommand \@ifnum [1]{%
 \ifnum #1\expandafter \@firstoftwo
 \else \expandafter \@secondoftwo
 \fi
}%
\providecommand \@ifx [1]{%
 \ifx #1\expandafter \@firstoftwo
 \else \expandafter \@secondoftwo
 \fi
}%
\providecommand \natexlab [1]{#1}%
\providecommand \enquote  [1]{``#1''}%
\providecommand \bibnamefont  [1]{#1}%
\providecommand \bibfnamefont [1]{#1}%
\providecommand \citenamefont [1]{#1}%
\providecommand \href@noop [0]{\@secondoftwo}%
\providecommand \href [0]{\begingroup \@sanitize@url \@href}%
\providecommand \@href[1]{\@@startlink{#1}\@@href}%
\providecommand \@@href[1]{\endgroup#1\@@endlink}%
\providecommand \@sanitize@url [0]{\catcode `\\12\catcode `\$12\catcode
  `\&12\catcode `\#12\catcode `\^12\catcode `\_12\catcode `\%12\relax}%
\providecommand \@@startlink[1]{}%
\providecommand \@@endlink[0]{}%
\providecommand \url  [0]{\begingroup\@sanitize@url \@url }%
\providecommand \@url [1]{\endgroup\@href {#1}{\urlprefix }}%
\providecommand \urlprefix  [0]{URL }%
\providecommand \Eprint [0]{\href }%
\providecommand \doibase [0]{http://dx.doi.org/}%
\providecommand \selectlanguage [0]{\@gobble}%
\providecommand \bibinfo  [0]{\@secondoftwo}%
\providecommand \bibfield  [0]{\@secondoftwo}%
\providecommand \translation [1]{[#1]}%
\providecommand \BibitemOpen [0]{}%
\providecommand \bibitemStop [0]{}%
\providecommand \bibitemNoStop [0]{.\EOS\space}%
\providecommand \EOS [0]{\spacefactor3000\relax}%
\providecommand \BibitemShut  [1]{\csname bibitem#1\endcsname}%
\let\auto@bib@innerbib\@empty
\bibitem [{\citenamefont {Rosenbluth}(1950)}]{Rosenbluth:1950yq}%
  \BibitemOpen
  \bibfield  {author} {\bibinfo {author} {\bibfnamefont {M.~N.}\ \bibnamefont
  {Rosenbluth}},\ }\href {\doibase 10.1103/PhysRev.79.615} {\bibfield
  {journal} {\bibinfo  {journal} {Phys.~Rev.}\ }\textbf {\bibinfo {volume}
  {79}},\ \bibinfo {pages} {615} (\bibinfo {year} {1950})}\BibitemShut
  {NoStop}%
\bibitem [{\citenamefont {Qattan}\ \emph {et~al.}(2005)\citenamefont {Qattan}
  \emph {et~al.}}]{Qattan:2004ht}%
  \BibitemOpen
  \bibfield  {author} {\bibinfo {author} {\bibfnamefont {I.~A.}\ \bibnamefont
  {Qattan}} \emph {et~al.},\ }\href {\doibase 10.1103/PhysRevLett.94.142301}
  {\bibfield  {journal} {\bibinfo  {journal} {Phys.~Rev.~Lett.}\ }\textbf
  {\bibinfo {volume} {94}},\ \bibinfo {pages} {142301} (\bibinfo {year}
  {2005})},\ \Eprint {http://arxiv.org/abs/nucl-ex/0410010}
  {arXiv:nucl-ex/0410010 [nucl-ex]} \BibitemShut {NoStop}%
\bibitem [{\citenamefont {Jones}\ \emph {et~al.}(2000)\citenamefont {Jones}
  \emph {et~al.}}]{Jones:1999rz}%
  \BibitemOpen
  \bibfield  {author} {\bibinfo {author} {\bibfnamefont {M.~K.}\ \bibnamefont
  {Jones}} \emph {et~al.} (\bibinfo {collaboration} {Jefferson Lab Hall A}),\
  }\href {\doibase 10.1103/PhysRevLett.84.1398} {\bibfield  {journal} {\bibinfo
   {journal} {Phys.~Rev.~Lett.}\ }\textbf {\bibinfo {volume} {84}},\ \bibinfo
  {pages} {1398} (\bibinfo {year} {2000})},\ \Eprint
  {http://arxiv.org/abs/nucl-ex/9910005} {arXiv:nucl-ex/9910005 [nucl-ex]}
  \BibitemShut {NoStop}%
\bibitem [{\citenamefont {Gayou}\ \emph {et~al.}(2002)\citenamefont {Gayou}
  \emph {et~al.}}]{Gayou:2001qd}%
  \BibitemOpen
  \bibfield  {author} {\bibinfo {author} {\bibfnamefont {O.}~\bibnamefont
  {Gayou}} \emph {et~al.} (\bibinfo {collaboration} {Jefferson Lab Hall A}),\
  }\href {\doibase 10.1103/PhysRevLett.88.092301} {\bibfield  {journal}
  {\bibinfo  {journal} {Phys.~Rev.~Lett.}\ }\textbf {\bibinfo {volume} {88}},\
  \bibinfo {pages} {092301} (\bibinfo {year} {2002})},\ \Eprint
  {http://arxiv.org/abs/nucl-ex/0111010} {arXiv:nucl-ex/0111010 [nucl-ex]}
  \BibitemShut {NoStop}%
\bibitem [{\citenamefont {Punjabi}\ \emph {et~al.}(2005)\citenamefont {Punjabi}
  \emph {et~al.}}]{Punjabi:2005wq}%
  \BibitemOpen
  \bibfield  {author} {\bibinfo {author} {\bibfnamefont {V.}~\bibnamefont
  {Punjabi}} \emph {et~al.},\ }\href {\doibase 10.1103/PhysRevC.71.055202,
  10.1103/PhysRevC.71.069902} {\bibfield  {journal} {\bibinfo  {journal}
  {Phys.~Rev.}\ }\textbf {\bibinfo {volume} {C71}},\ \bibinfo {pages} {055202}
  (\bibinfo {year} {2005})},\ \bibinfo {note} {[Erratum: Phys.
  Rev.C71,069902(2005)]},\ \Eprint {http://arxiv.org/abs/nucl-ex/0501018}
  {arXiv:nucl-ex/0501018 [nucl-ex]} \BibitemShut {NoStop}%
\bibitem [{\citenamefont {Puckett}\ \emph {et~al.}(2010)\citenamefont {Puckett}
  \emph {et~al.}}]{Puckett:2010ac}%
  \BibitemOpen
  \bibfield  {author} {\bibinfo {author} {\bibfnamefont {A.~J.~R.}\
  \bibnamefont {Puckett}} \emph {et~al.},\ }\href {\doibase
  10.1103/PhysRevLett.104.242301} {\bibfield  {journal} {\bibinfo  {journal}
  {Phys.~Rev.~Lett.}\ }\textbf {\bibinfo {volume} {104}},\ \bibinfo {pages}
  {242301} (\bibinfo {year} {2010})},\ \Eprint {http://arxiv.org/abs/1005.3419}
  {arXiv:1005.3419 [nucl-ex]} \BibitemShut {NoStop}%
\bibitem [{\citenamefont {Puckett}\ \emph {et~al.}(2012)\citenamefont {Puckett}
  \emph {et~al.}}]{Puckett:2011xg}%
  \BibitemOpen
  \bibfield  {author} {\bibinfo {author} {\bibfnamefont {A.~J.~R.}\
  \bibnamefont {Puckett}} \emph {et~al.},\ }\href {\doibase
  10.1103/PhysRevC.85.045203} {\bibfield  {journal} {\bibinfo  {journal}
  {Phys.~Rev.}\ }\textbf {\bibinfo {volume} {C85}},\ \bibinfo {pages} {045203}
  (\bibinfo {year} {2012})},\ \Eprint {http://arxiv.org/abs/1102.5737}
  {arXiv:1102.5737 [nucl-ex]} \BibitemShut {NoStop}%
\bibitem [{\citenamefont {Guichon}\ and\ \citenamefont
  {Vanderhaeghen}(2003)}]{Guichon:2003qm}%
  \BibitemOpen
  \bibfield  {author} {\bibinfo {author} {\bibfnamefont {P.~A.~M.}\
  \bibnamefont {Guichon}}\ and\ \bibinfo {author} {\bibfnamefont
  {M.}~\bibnamefont {Vanderhaeghen}},\ }\href {\doibase
  10.1103/PhysRevLett.91.142303} {\bibfield  {journal} {\bibinfo  {journal}
  {Phys.~Rev.~Lett.}\ }\textbf {\bibinfo {volume} {91}},\ \bibinfo {pages}
  {142303} (\bibinfo {year} {2003})},\ \Eprint
  {http://arxiv.org/abs/hep-ph/0306007} {arXiv:hep-ph/0306007 [hep-ph]}
  \BibitemShut {NoStop}%
\bibitem [{\citenamefont {Blunden}\ \emph {et~al.}(2005)\citenamefont
  {Blunden}, \citenamefont {Melnitchouk},\ and\ \citenamefont
  {Tjon}}]{Blunden:2005ew}%
  \BibitemOpen
  \bibfield  {author} {\bibinfo {author} {\bibfnamefont {P.~G.}\ \bibnamefont
  {Blunden}}, \bibinfo {author} {\bibfnamefont {W.}~\bibnamefont
  {Melnitchouk}}, \ and\ \bibinfo {author} {\bibfnamefont {J.~A.}\ \bibnamefont
  {Tjon}},\ }\href {\doibase 10.1103/PhysRevC.72.034612} {\bibfield  {journal}
  {\bibinfo  {journal} {Phys.~Rev.}\ }\textbf {\bibinfo {volume} {C72}},\
  \bibinfo {pages} {034612} (\bibinfo {year} {2005})},\ \Eprint
  {http://arxiv.org/abs/nucl-th/0506039} {arXiv:nucl-th/0506039 [nucl-th]}
  \BibitemShut {NoStop}%
\bibitem [{\citenamefont {Volmer}\ \emph {et~al.}(2001)\citenamefont {Volmer}
  \emph {et~al.}}]{Volmer:2000ek}%
  \BibitemOpen
  \bibfield  {author} {\bibinfo {author} {\bibfnamefont {J.}~\bibnamefont
  {Volmer}} \emph {et~al.} (\bibinfo {collaboration} {Jefferson Lab F(pi)}),\
  }\href {\doibase 10.1103/PhysRevLett.86.1713} {\bibfield  {journal} {\bibinfo
   {journal} {Phys.~Rev.~Lett.}\ }\textbf {\bibinfo {volume} {86}},\ \bibinfo
  {pages} {1713} (\bibinfo {year} {2001})},\ \Eprint
  {http://arxiv.org/abs/nucl-ex/0010009} {arXiv:nucl-ex/0010009 [nucl-ex]}
  \BibitemShut {NoStop}%
\bibitem [{\citenamefont {Horn}\ \emph {et~al.}(2006)\citenamefont {Horn} \emph
  {et~al.}}]{Horn:2006tm}%
  \BibitemOpen
  \bibfield  {author} {\bibinfo {author} {\bibfnamefont {T.}~\bibnamefont
  {Horn}} \emph {et~al.} (\bibinfo {collaboration} {Jefferson Lab F(pi)-2}),\
  }\href {\doibase 10.1103/PhysRevLett.97.192001} {\bibfield  {journal}
  {\bibinfo  {journal} {Phys.~Rev.~Lett.}\ }\textbf {\bibinfo {volume} {97}},\
  \bibinfo {pages} {192001} (\bibinfo {year} {2006})},\ \Eprint
  {http://arxiv.org/abs/nucl-ex/0607005} {arXiv:nucl-ex/0607005 [nucl-ex]}
  \BibitemShut {NoStop}%
\bibitem [{\citenamefont {Huber}\ \emph {et~al.}(2008)\citenamefont {Huber}
  \emph {et~al.}}]{Huber:2008id}%
  \BibitemOpen
  \bibfield  {author} {\bibinfo {author} {\bibfnamefont {G.~M.}\ \bibnamefont
  {Huber}} \emph {et~al.} (\bibinfo {collaboration} {Jefferson Lab}),\ }\href
  {\doibase 10.1103/PhysRevC.78.045203} {\bibfield  {journal} {\bibinfo
  {journal} {Phys.~Rev.}\ }\textbf {\bibinfo {volume} {C78}},\ \bibinfo {pages}
  {045203} (\bibinfo {year} {2008})},\ \Eprint {http://arxiv.org/abs/0809.3052}
  {arXiv:0809.3052 [nucl-ex]} \BibitemShut {NoStop}%
\bibitem [{\citenamefont {Chang}\ \emph {et~al.}(2013)\citenamefont {Chang}
  \emph {et~al.}}]{Chang:2013nia}%
  \BibitemOpen
  \bibfield  {author} {\bibinfo {author} {\bibfnamefont {L.}~\bibnamefont
  {Chang}} \emph {et~al.},\ }\href {\doibase 10.1103/PhysRevLett.111.141802}
  {\bibfield  {journal} {\bibinfo  {journal} {Phys.~Rev.~Lett.}\ }\textbf
  {\bibinfo {volume} {111}},\ \bibinfo {pages} {141802} (\bibinfo {year}
  {2013})},\ \Eprint {http://arxiv.org/abs/1307.0026} {arXiv:1307.0026
  [nucl-th]} \BibitemShut {NoStop}%
\bibitem [{\citenamefont {Horn}\ and\ \citenamefont
  {Roberts}(2016)}]{Horn:2016rip}%
  \BibitemOpen
  \bibfield  {author} {\bibinfo {author} {\bibfnamefont {T.}~\bibnamefont
  {Horn}}\ and\ \bibinfo {author} {\bibfnamefont {C.~D.}\ \bibnamefont
  {Roberts}},\ }\href {\doibase 10.1088/0954-3899/43/7/073001} {\bibfield
  {journal} {\bibinfo  {journal} {J.~Phys.}\ }\textbf {\bibinfo {volume}
  {G43}},\ \bibinfo {pages} {073001} (\bibinfo {year} {2016})},\ \Eprint
  {http://arxiv.org/abs/1602.04016} {arXiv:1602.04016 [nucl-th]} \BibitemShut
  {NoStop}%
\bibitem [{\citenamefont {Collins}\ \emph {et~al.}(2011)\citenamefont {Collins}
  \emph {et~al.}}]{Collins:2011mk}%
  \BibitemOpen
  \bibfield  {author} {\bibinfo {author} {\bibfnamefont {S.}~\bibnamefont
  {Collins}} \emph {et~al.},\ }\href {\doibase 10.1103/PhysRevD.84.074507}
  {\bibfield  {journal} {\bibinfo  {journal} {Phys.~Rev.}\ }\textbf {\bibinfo
  {volume} {D84}},\ \bibinfo {pages} {074507} (\bibinfo {year} {2011})},\
  \Eprint {http://arxiv.org/abs/1106.3580} {arXiv:1106.3580 [hep-lat]}
  \BibitemShut {NoStop}%
\bibitem [{\citenamefont {Alexandrou}\ \emph {et~al.}(2011)\citenamefont
  {Alexandrou}, \citenamefont {Brinet}, \citenamefont {Carbonell},
  \citenamefont {Constantinou}, \citenamefont {Harraud}, \citenamefont
  {Guichon}, \citenamefont {Jansen}, \citenamefont {Korzec},\ and\
  \citenamefont {Papinutto}}]{Alexandrou:2011db}%
  \BibitemOpen
  \bibfield  {author} {\bibinfo {author} {\bibfnamefont {C.}~\bibnamefont
  {Alexandrou}}, \bibinfo {author} {\bibfnamefont {M.}~\bibnamefont {Brinet}},
  \bibinfo {author} {\bibfnamefont {J.}~\bibnamefont {Carbonell}}, \bibinfo
  {author} {\bibfnamefont {M.}~\bibnamefont {Constantinou}}, \bibinfo {author}
  {\bibfnamefont {P.~A.}\ \bibnamefont {Harraud}}, \bibinfo {author}
  {\bibfnamefont {P.}~\bibnamefont {Guichon}}, \bibinfo {author} {\bibfnamefont
  {K.}~\bibnamefont {Jansen}}, \bibinfo {author} {\bibfnamefont
  {T.}~\bibnamefont {Korzec}}, \ and\ \bibinfo {author} {\bibfnamefont
  {M.}~\bibnamefont {Papinutto}},\ }\href {\doibase 10.1103/PhysRevD.83.094502}
  {\bibfield  {journal} {\bibinfo  {journal} {Phys.~Rev.}\ }\textbf {\bibinfo
  {volume} {D83}},\ \bibinfo {pages} {094502} (\bibinfo {year} {2011})},\
  \Eprint {http://arxiv.org/abs/1102.2208} {arXiv:1102.2208 [hep-lat]}
  \BibitemShut {NoStop}%
\bibitem [{\citenamefont {Shanahan}\ \emph
  {et~al.}(2014{\natexlab{a}})\citenamefont {Shanahan}, \citenamefont {Thomas},
  \citenamefont {Young}, \citenamefont {Zanotti}, \citenamefont {Horsley},
  \citenamefont {Nakamura}, \citenamefont {Pleiter}, \citenamefont {Rakow},
  \citenamefont {Schierholz},\ and\ \citenamefont
  {Stüben}}]{Shanahan:2014uka}%
  \BibitemOpen
  \bibfield  {author} {\bibinfo {author} {\bibfnamefont {P.~E.}\ \bibnamefont
  {Shanahan}}, \bibinfo {author} {\bibfnamefont {A.~W.}\ \bibnamefont
  {Thomas}}, \bibinfo {author} {\bibfnamefont {R.~D.}\ \bibnamefont {Young}},
  \bibinfo {author} {\bibfnamefont {J.~M.}\ \bibnamefont {Zanotti}}, \bibinfo
  {author} {\bibfnamefont {R.}~\bibnamefont {Horsley}}, \bibinfo {author}
  {\bibfnamefont {Y.}~\bibnamefont {Nakamura}}, \bibinfo {author}
  {\bibfnamefont {D.}~\bibnamefont {Pleiter}}, \bibinfo {author} {\bibfnamefont
  {P.~E.~L.}\ \bibnamefont {Rakow}}, \bibinfo {author} {\bibfnamefont
  {G.}~\bibnamefont {Schierholz}}, \ and\ \bibinfo {author} {\bibfnamefont
  {H.}~\bibnamefont {Stüben}} (\bibinfo {collaboration} {QCDSF/UKQCD, CSSM}),\
  }\href {\doibase 10.1103/PhysRevD.89.074511} {\bibfield  {journal} {\bibinfo
  {journal} {Phys.~Rev.}\ }\textbf {\bibinfo {volume} {D89}},\ \bibinfo {pages}
  {074511} (\bibinfo {year} {2014}{\natexlab{a}})},\ \Eprint
  {http://arxiv.org/abs/1401.5862} {arXiv:1401.5862 [hep-lat]} \BibitemShut
  {NoStop}%
\bibitem [{\citenamefont {Shanahan}\ \emph
  {et~al.}(2014{\natexlab{b}})\citenamefont {Shanahan}, \citenamefont {Thomas},
  \citenamefont {Young}, \citenamefont {Zanotti}, \citenamefont {Horsley},
  \citenamefont {Nakamura}, \citenamefont {Pleiter}, \citenamefont {Rakow},
  \citenamefont {Schierholz},\ and\ \citenamefont
  {Stüben}}]{Shanahan:2014cga}%
  \BibitemOpen
  \bibfield  {author} {\bibinfo {author} {\bibfnamefont {P.~E.}\ \bibnamefont
  {Shanahan}}, \bibinfo {author} {\bibfnamefont {A.~W.}\ \bibnamefont
  {Thomas}}, \bibinfo {author} {\bibfnamefont {R.~D.}\ \bibnamefont {Young}},
  \bibinfo {author} {\bibfnamefont {J.~M.}\ \bibnamefont {Zanotti}}, \bibinfo
  {author} {\bibfnamefont {R.}~\bibnamefont {Horsley}}, \bibinfo {author}
  {\bibfnamefont {Y.}~\bibnamefont {Nakamura}}, \bibinfo {author}
  {\bibfnamefont {D.}~\bibnamefont {Pleiter}}, \bibinfo {author} {\bibfnamefont
  {P.~E.~L.}\ \bibnamefont {Rakow}}, \bibinfo {author} {\bibfnamefont
  {G.}~\bibnamefont {Schierholz}}, \ and\ \bibinfo {author} {\bibfnamefont
  {H.}~\bibnamefont {Stüben}},\ }\href {\doibase 10.1103/PhysRevD.90.034502}
  {\bibfield  {journal} {\bibinfo  {journal} {Phys.~Rev.}\ }\textbf {\bibinfo
  {volume} {D90}},\ \bibinfo {pages} {034502} (\bibinfo {year}
  {2014}{\natexlab{b}})},\ \Eprint {http://arxiv.org/abs/1403.1965}
  {arXiv:1403.1965 [hep-lat]} \BibitemShut {NoStop}%
\bibitem [{\citenamefont {Green}\ \emph {et~al.}(2014)\citenamefont {Green},
  \citenamefont {Negele}, \citenamefont {Pochinsky}, \citenamefont {Syritsyn},
  \citenamefont {Engelhardt},\ and\ \citenamefont {Krieg}}]{Green:2014xba}%
  \BibitemOpen
  \bibfield  {author} {\bibinfo {author} {\bibfnamefont {J.~R.}\ \bibnamefont
  {Green}}, \bibinfo {author} {\bibfnamefont {J.~W.}\ \bibnamefont {Negele}},
  \bibinfo {author} {\bibfnamefont {A.~V.}\ \bibnamefont {Pochinsky}}, \bibinfo
  {author} {\bibfnamefont {S.~N.}\ \bibnamefont {Syritsyn}}, \bibinfo {author}
  {\bibfnamefont {M.}~\bibnamefont {Engelhardt}}, \ and\ \bibinfo {author}
  {\bibfnamefont {S.}~\bibnamefont {Krieg}},\ }\href {\doibase
  10.1103/PhysRevD.90.074507} {\bibfield  {journal} {\bibinfo  {journal}
  {Phys.~Rev.}\ }\textbf {\bibinfo {volume} {D90}},\ \bibinfo {pages} {074507}
  (\bibinfo {year} {2014})},\ \Eprint {http://arxiv.org/abs/1404.4029}
  {arXiv:1404.4029 [hep-lat]} \BibitemShut {NoStop}%
\bibitem [{\citenamefont {Capitani}\ \emph {et~al.}(2015)\citenamefont
  {Capitani}, \citenamefont {Della~Morte}, \citenamefont {Djukanovic},
  \citenamefont {von Hippel}, \citenamefont {Hua}, \citenamefont {Jäger},
  \citenamefont {Knippschild}, \citenamefont {Meyer}, \citenamefont {Rae},\
  and\ \citenamefont {Wittig}}]{Capitani:2015sba}%
  \BibitemOpen
  \bibfield  {author} {\bibinfo {author} {\bibfnamefont {S.}~\bibnamefont
  {Capitani}}, \bibinfo {author} {\bibfnamefont {M.}~\bibnamefont
  {Della~Morte}}, \bibinfo {author} {\bibfnamefont {D.}~\bibnamefont
  {Djukanovic}}, \bibinfo {author} {\bibfnamefont {G.}~\bibnamefont {von
  Hippel}}, \bibinfo {author} {\bibfnamefont {J.}~\bibnamefont {Hua}}, \bibinfo
  {author} {\bibfnamefont {B.}~\bibnamefont {Jäger}}, \bibinfo {author}
  {\bibfnamefont {B.}~\bibnamefont {Knippschild}}, \bibinfo {author}
  {\bibfnamefont {H.~B.}\ \bibnamefont {Meyer}}, \bibinfo {author}
  {\bibfnamefont {T.~D.}\ \bibnamefont {Rae}}, \ and\ \bibinfo {author}
  {\bibfnamefont {H.}~\bibnamefont {Wittig}},\ }\href {\doibase
  10.1103/PhysRevD.92.054511} {\bibfield  {journal} {\bibinfo  {journal}
  {Phys.~Rev.}\ }\textbf {\bibinfo {volume} {D92}},\ \bibinfo {pages} {054511}
  (\bibinfo {year} {2015})},\ \Eprint {http://arxiv.org/abs/1504.04628}
  {arXiv:1504.04628 [hep-lat]} \BibitemShut {NoStop}%
\bibitem [{\citenamefont {Lin}\ \emph {et~al.}(2010)\citenamefont {Lin},
  \citenamefont {Cohen}, \citenamefont {Edwards}, \citenamefont {Orginos},\
  and\ \citenamefont {Richards}}]{Lin:2010fv}%
  \BibitemOpen
  \bibfield  {author} {\bibinfo {author} {\bibfnamefont {H.-W.}\ \bibnamefont
  {Lin}}, \bibinfo {author} {\bibfnamefont {S.~D.}\ \bibnamefont {Cohen}},
  \bibinfo {author} {\bibfnamefont {R.~G.}\ \bibnamefont {Edwards}}, \bibinfo
  {author} {\bibfnamefont {K.}~\bibnamefont {Orginos}}, \ and\ \bibinfo
  {author} {\bibfnamefont {D.~G.}\ \bibnamefont {Richards}},\ }\href@noop {} {\
   (\bibinfo {year} {2010})},\ \Eprint {http://arxiv.org/abs/1005.0799}
  {arXiv:1005.0799 [hep-lat]} \BibitemShut {NoStop}%
\bibitem [{\citenamefont {Koponen}\ \emph {et~al.}(2017)\citenamefont
  {Koponen}, \citenamefont {Zimermmane-Santos}, \citenamefont {Davies},
  \citenamefont {Lepage},\ and\ \citenamefont {Lytle}}]{Koponen:2017fvm}%
  \BibitemOpen
  \bibfield  {author} {\bibinfo {author} {\bibfnamefont {J.}~\bibnamefont
  {Koponen}}, \bibinfo {author} {\bibfnamefont {A.~C.}\ \bibnamefont
  {Zimermmane-Santos}}, \bibinfo {author} {\bibfnamefont {C.~T.~H.}\
  \bibnamefont {Davies}}, \bibinfo {author} {\bibfnamefont {G.~P.}\
  \bibnamefont {Lepage}}, \ and\ \bibinfo {author} {\bibfnamefont {A.~T.}\
  \bibnamefont {Lytle}},\ }\href@noop {} {\  (\bibinfo {year} {2017})},\
  \Eprint {http://arxiv.org/abs/1701.04250} {arXiv:1701.04250 [hep-lat]}
  \BibitemShut {NoStop}%
\bibitem [{\citenamefont {Owen}\ \emph {et~al.}(2013)\citenamefont {Owen},
  \citenamefont {Dragos}, \citenamefont {Kamleh}, \citenamefont {Leinweber},
  \citenamefont {Mahbub}, \citenamefont {Menadue},\ and\ \citenamefont
  {Zanotti}}]{Owen:2012ts}%
  \BibitemOpen
  \bibfield  {author} {\bibinfo {author} {\bibfnamefont {B.~J.}\ \bibnamefont
  {Owen}}, \bibinfo {author} {\bibfnamefont {J.}~\bibnamefont {Dragos}},
  \bibinfo {author} {\bibfnamefont {W.}~\bibnamefont {Kamleh}}, \bibinfo
  {author} {\bibfnamefont {D.~B.}\ \bibnamefont {Leinweber}}, \bibinfo {author}
  {\bibfnamefont {M.~S.}\ \bibnamefont {Mahbub}}, \bibinfo {author}
  {\bibfnamefont {B.~J.}\ \bibnamefont {Menadue}}, \ and\ \bibinfo {author}
  {\bibfnamefont {J.~M.}\ \bibnamefont {Zanotti}},\ }\href {\doibase
  10.1016/j.physletb.2013.04.063} {\bibfield  {journal} {\bibinfo  {journal}
  {Phys.~Lett.}\ }\textbf {\bibinfo {volume} {B723}},\ \bibinfo {pages} {217}
  (\bibinfo {year} {2013})},\ \Eprint {http://arxiv.org/abs/1212.4668}
  {arXiv:1212.4668 [hep-lat]} \BibitemShut {NoStop}%
\bibitem [{\citenamefont {Yoon}\ \emph {et~al.}(2016)\citenamefont {Yoon} \emph
  {et~al.}}]{Yoon:2016dij}%
  \BibitemOpen
  \bibfield  {author} {\bibinfo {author} {\bibfnamefont {B.}~\bibnamefont
  {Yoon}} \emph {et~al.},\ }\href {\doibase 10.1103/PhysRevD.93.114506}
  {\bibfield  {journal} {\bibinfo  {journal} {Phys.~Rev.}\ }\textbf {\bibinfo
  {volume} {D93}},\ \bibinfo {pages} {114506} (\bibinfo {year} {2016})},\
  \Eprint {http://arxiv.org/abs/1602.07737} {arXiv:1602.07737 [hep-lat]}
  \BibitemShut {NoStop}%
\bibitem [{\citenamefont {Dragos}\ \emph {et~al.}(2016)\citenamefont {Dragos},
  \citenamefont {Horsley}, \citenamefont {Kamleh}, \citenamefont {Leinweber},
  \citenamefont {Nakamura}, \citenamefont {Rakow}, \citenamefont {Schierholz},
  \citenamefont {Young},\ and\ \citenamefont {Zanotti}}]{Dragos:2016rtx}%
  \BibitemOpen
  \bibfield  {author} {\bibinfo {author} {\bibfnamefont {J.}~\bibnamefont
  {Dragos}}, \bibinfo {author} {\bibfnamefont {R.}~\bibnamefont {Horsley}},
  \bibinfo {author} {\bibfnamefont {W.}~\bibnamefont {Kamleh}}, \bibinfo
  {author} {\bibfnamefont {D.~B.}\ \bibnamefont {Leinweber}}, \bibinfo {author}
  {\bibfnamefont {Y.}~\bibnamefont {Nakamura}}, \bibinfo {author}
  {\bibfnamefont {P.~E.~L.}\ \bibnamefont {Rakow}}, \bibinfo {author}
  {\bibfnamefont {G.}~\bibnamefont {Schierholz}}, \bibinfo {author}
  {\bibfnamefont {R.~D.}\ \bibnamefont {Young}}, \ and\ \bibinfo {author}
  {\bibfnamefont {J.~M.}\ \bibnamefont {Zanotti}},\ }\href {\doibase
  10.1103/PhysRevD.94.074505} {\bibfield  {journal} {\bibinfo  {journal}
  {Phys.~Rev.}\ }\textbf {\bibinfo {volume} {D94}},\ \bibinfo {pages} {074505}
  (\bibinfo {year} {2016})},\ \Eprint {http://arxiv.org/abs/1606.03195}
  {arXiv:1606.03195 [hep-lat]} \BibitemShut {NoStop}%
\bibitem [{\citenamefont {Horsley}\ \emph {et~al.}(2012)\citenamefont
  {Horsley}, \citenamefont {Millo}, \citenamefont {Nakamura}, \citenamefont
  {Perlt}, \citenamefont {Pleiter}, \citenamefont {Rakow}, \citenamefont
  {Schierholz}, \citenamefont {Schiller}, \citenamefont {Winter},\ and\
  \citenamefont {Zanotti}}]{Horsley:2012pz}%
  \BibitemOpen
  \bibfield  {author} {\bibinfo {author} {\bibfnamefont {R.}~\bibnamefont
  {Horsley}}, \bibinfo {author} {\bibfnamefont {R.}~\bibnamefont {Millo}},
  \bibinfo {author} {\bibfnamefont {Y.}~\bibnamefont {Nakamura}}, \bibinfo
  {author} {\bibfnamefont {H.}~\bibnamefont {Perlt}}, \bibinfo {author}
  {\bibfnamefont {D.}~\bibnamefont {Pleiter}}, \bibinfo {author} {\bibfnamefont
  {P.~E.~L.}\ \bibnamefont {Rakow}}, \bibinfo {author} {\bibfnamefont
  {G.}~\bibnamefont {Schierholz}}, \bibinfo {author} {\bibfnamefont
  {A.}~\bibnamefont {Schiller}}, \bibinfo {author} {\bibfnamefont
  {F.}~\bibnamefont {Winter}}, \ and\ \bibinfo {author} {\bibfnamefont {J.~M.}\
  \bibnamefont {Zanotti}} (\bibinfo {collaboration} {UKQCD, QCDSF}),\ }\href
  {\doibase 10.1016/j.physletb.2012.07.004} {\bibfield  {journal} {\bibinfo
  {journal} {Phys.~Lett.}\ }\textbf {\bibinfo {volume} {B714}},\ \bibinfo
  {pages} {312} (\bibinfo {year} {2012})},\ \Eprint
  {http://arxiv.org/abs/1205.6410} {arXiv:1205.6410 [hep-lat]} \BibitemShut
  {NoStop}%
\bibitem [{\citenamefont {Chambers}\ \emph {et~al.}(2014)\citenamefont
  {Chambers} \emph {et~al.}}]{Chambers:2014qaa}%
  \BibitemOpen
  \bibfield  {author} {\bibinfo {author} {\bibfnamefont {A.~J.}\ \bibnamefont
  {Chambers}} \emph {et~al.} (\bibinfo {collaboration} {QCDSF/UKQCD, CSSM}),\
  }\href {\doibase 10.1103/PhysRevD.90.014510} {\bibfield  {journal} {\bibinfo
  {journal} {Phys.~Rev.}\ }\textbf {\bibinfo {volume} {D90}},\ \bibinfo {pages}
  {014510} (\bibinfo {year} {2014})},\ \Eprint {http://arxiv.org/abs/1405.3019}
  {arXiv:1405.3019 [hep-lat]} \BibitemShut {NoStop}%
\bibitem [{\citenamefont {Chambers}\ \emph
  {et~al.}(2015{\natexlab{a}})\citenamefont {Chambers}, \citenamefont
  {Horsley}, \citenamefont {Nakamura}, \citenamefont {Perlt}, \citenamefont
  {Rakow}, \citenamefont {Schierholz}, \citenamefont {Schiller},\ and\
  \citenamefont {Zanotti}}]{Chambers:2014pea}%
  \BibitemOpen
  \bibfield  {author} {\bibinfo {author} {\bibfnamefont {A.~J.}\ \bibnamefont
  {Chambers}}, \bibinfo {author} {\bibfnamefont {R.}~\bibnamefont {Horsley}},
  \bibinfo {author} {\bibfnamefont {Y.}~\bibnamefont {Nakamura}}, \bibinfo
  {author} {\bibfnamefont {H.}~\bibnamefont {Perlt}}, \bibinfo {author}
  {\bibfnamefont {P.~E.~L.}\ \bibnamefont {Rakow}}, \bibinfo {author}
  {\bibfnamefont {G.}~\bibnamefont {Schierholz}}, \bibinfo {author}
  {\bibfnamefont {A.}~\bibnamefont {Schiller}}, \ and\ \bibinfo {author}
  {\bibfnamefont {J.~M.}\ \bibnamefont {Zanotti}} (\bibinfo {collaboration}
  {QCDSF}),\ }\href {\doibase 10.1016/j.physletb.2014.11.033} {\bibfield
  {journal} {\bibinfo  {journal} {Phys.~Lett.}\ }\textbf {\bibinfo {volume}
  {B740}},\ \bibinfo {pages} {30} (\bibinfo {year} {2015}{\natexlab{a}})},\
  \Eprint {http://arxiv.org/abs/1410.3078} {arXiv:1410.3078 [hep-lat]}
  \BibitemShut {NoStop}%
\bibitem [{\citenamefont {Chambers}\ \emph
  {et~al.}(2015{\natexlab{b}})\citenamefont {Chambers} \emph
  {et~al.}}]{Chambers:2015bka}%
  \BibitemOpen
  \bibfield  {author} {\bibinfo {author} {\bibfnamefont {A.~J.}\ \bibnamefont
  {Chambers}} \emph {et~al.},\ }\href {\doibase 10.1103/PhysRevD.92.114517}
  {\bibfield  {journal} {\bibinfo  {journal} {Phys.~Rev.}\ }\textbf {\bibinfo
  {volume} {D92}},\ \bibinfo {pages} {114517} (\bibinfo {year}
  {2015}{\natexlab{b}})},\ \Eprint {http://arxiv.org/abs/1508.06856}
  {arXiv:1508.06856 [hep-lat]} \BibitemShut {NoStop}%
\bibitem [{\citenamefont {Detmold}(2005)}]{Detmold:2004kw}%
  \BibitemOpen
  \bibfield  {author} {\bibinfo {author} {\bibfnamefont {W.}~\bibnamefont
  {Detmold}},\ }\href {\doibase 10.1103/PhysRevD.71.054506} {\bibfield
  {journal} {\bibinfo  {journal} {Phys.~Rev.}\ }\textbf {\bibinfo {volume}
  {D71}},\ \bibinfo {pages} {054506} (\bibinfo {year} {2005})},\ \Eprint
  {http://arxiv.org/abs/hep-lat/0410011} {arXiv:hep-lat/0410011 [hep-lat]}
  \BibitemShut {NoStop}%
\bibitem [{\citenamefont {Engelhardt}(2007)}]{Engelhardt:2007ub}%
  \BibitemOpen
  \bibfield  {author} {\bibinfo {author} {\bibfnamefont {M.}~\bibnamefont
  {Engelhardt}} (\bibinfo {collaboration} {LHPC}),\ }\href {\doibase
  10.1103/PhysRevD.76.114502} {\bibfield  {journal} {\bibinfo  {journal}
  {Phys.~Rev.}\ }\textbf {\bibinfo {volume} {D76}},\ \bibinfo {pages} {114502}
  (\bibinfo {year} {2007})},\ \Eprint {http://arxiv.org/abs/0706.3919}
  {arXiv:0706.3919 [hep-lat]} \BibitemShut {NoStop}%
\bibitem [{\citenamefont {Detmold}\ \emph {et~al.}(2009)\citenamefont
  {Detmold}, \citenamefont {Tiburzi},\ and\ \citenamefont
  {Walker-Loud}}]{Detmold:2009dx}%
  \BibitemOpen
  \bibfield  {author} {\bibinfo {author} {\bibfnamefont {W.}~\bibnamefont
  {Detmold}}, \bibinfo {author} {\bibfnamefont {B.~C.}\ \bibnamefont
  {Tiburzi}}, \ and\ \bibinfo {author} {\bibfnamefont {A.}~\bibnamefont
  {Walker-Loud}},\ }\href {\doibase 10.1103/PhysRevD.79.094505} {\bibfield
  {journal} {\bibinfo  {journal} {Phys.~Rev.}\ }\textbf {\bibinfo {volume}
  {D79}},\ \bibinfo {pages} {094505} (\bibinfo {year} {2009})},\ \Eprint
  {http://arxiv.org/abs/0904.1586} {arXiv:0904.1586 [hep-lat]} \BibitemShut
  {NoStop}%
\bibitem [{\citenamefont {Detmold}\ \emph {et~al.}(2010)\citenamefont
  {Detmold}, \citenamefont {Tiburzi},\ and\ \citenamefont
  {Walker-Loud}}]{Detmold:2010ts}%
  \BibitemOpen
  \bibfield  {author} {\bibinfo {author} {\bibfnamefont {W.}~\bibnamefont
  {Detmold}}, \bibinfo {author} {\bibfnamefont {B.~C.}\ \bibnamefont
  {Tiburzi}}, \ and\ \bibinfo {author} {\bibfnamefont {A.}~\bibnamefont
  {Walker-Loud}},\ }\href {\doibase 10.1103/PhysRevD.81.054502} {\bibfield
  {journal} {\bibinfo  {journal} {Phys.~Rev.}\ }\textbf {\bibinfo {volume}
  {D81}},\ \bibinfo {pages} {054502} (\bibinfo {year} {2010})},\ \Eprint
  {http://arxiv.org/abs/1001.1131} {arXiv:1001.1131 [hep-lat]} \BibitemShut
  {NoStop}%
\bibitem [{\citenamefont {Primer}\ \emph {et~al.}(2014)\citenamefont {Primer},
  \citenamefont {Kamleh}, \citenamefont {Leinweber},\ and\ \citenamefont
  {Burkardt}}]{Primer:2013pva}%
  \BibitemOpen
  \bibfield  {author} {\bibinfo {author} {\bibfnamefont {T.}~\bibnamefont
  {Primer}}, \bibinfo {author} {\bibfnamefont {W.}~\bibnamefont {Kamleh}},
  \bibinfo {author} {\bibfnamefont {D.}~\bibnamefont {Leinweber}}, \ and\
  \bibinfo {author} {\bibfnamefont {M.}~\bibnamefont {Burkardt}},\ }\href
  {\doibase 10.1103/PhysRevD.89.034508} {\bibfield  {journal} {\bibinfo
  {journal} {Phys.~Rev.}\ }\textbf {\bibinfo {volume} {D89}},\ \bibinfo {pages}
  {034508} (\bibinfo {year} {2014})},\ \Eprint {http://arxiv.org/abs/1307.1509}
  {arXiv:1307.1509 [hep-lat]} \BibitemShut {NoStop}%
\bibitem [{\citenamefont {Freeman}\ \emph {et~al.}(2014)\citenamefont
  {Freeman}, \citenamefont {Alexandru}, \citenamefont {Lujan},\ and\
  \citenamefont {Lee}}]{Freeman:2014kka}%
  \BibitemOpen
  \bibfield  {author} {\bibinfo {author} {\bibfnamefont {W.}~\bibnamefont
  {Freeman}}, \bibinfo {author} {\bibfnamefont {A.}~\bibnamefont {Alexandru}},
  \bibinfo {author} {\bibfnamefont {M.}~\bibnamefont {Lujan}}, \ and\ \bibinfo
  {author} {\bibfnamefont {F.~X.}\ \bibnamefont {Lee}},\ }\href {\doibase
  10.1103/PhysRevD.90.054507} {\bibfield  {journal} {\bibinfo  {journal}
  {Phys.~Rev.}\ }\textbf {\bibinfo {volume} {D90}},\ \bibinfo {pages} {054507}
  (\bibinfo {year} {2014})},\ \Eprint {http://arxiv.org/abs/1407.2687}
  {arXiv:1407.2687 [hep-lat]} \BibitemShut {NoStop}%
\bibitem [{\citenamefont {Savage}\ \emph {et~al.}(2016)\citenamefont {Savage},
  \citenamefont {Shanahan}, \citenamefont {Tiburzi}, \citenamefont {Wagman},
  \citenamefont {Winter}, \citenamefont {Beane}, \citenamefont {Chang},
  \citenamefont {Davoudi}, \citenamefont {Detmold},\ and\ \citenamefont
  {Orginos}}]{Savage:2016kon}%
  \BibitemOpen
  \bibfield  {author} {\bibinfo {author} {\bibfnamefont {M.~J.}\ \bibnamefont
  {Savage}}, \bibinfo {author} {\bibfnamefont {P.~E.}\ \bibnamefont
  {Shanahan}}, \bibinfo {author} {\bibfnamefont {B.~C.}\ \bibnamefont
  {Tiburzi}}, \bibinfo {author} {\bibfnamefont {M.~L.}\ \bibnamefont {Wagman}},
  \bibinfo {author} {\bibfnamefont {F.}~\bibnamefont {Winter}}, \bibinfo
  {author} {\bibfnamefont {S.~R.}\ \bibnamefont {Beane}}, \bibinfo {author}
  {\bibfnamefont {E.}~\bibnamefont {Chang}}, \bibinfo {author} {\bibfnamefont
  {Z.}~\bibnamefont {Davoudi}}, \bibinfo {author} {\bibfnamefont
  {W.}~\bibnamefont {Detmold}}, \ and\ \bibinfo {author} {\bibfnamefont
  {K.}~\bibnamefont {Orginos}},\ }\href@noop {} {\  (\bibinfo {year} {2016})},\
  \Eprint {http://arxiv.org/abs/1610.04545} {arXiv:1610.04545 [hep-lat]}
  \BibitemShut {NoStop}%
\bibitem [{\citenamefont {Bouchard}\ \emph {et~al.}(2016)\citenamefont
  {Bouchard}, \citenamefont {Chang}, \citenamefont {Kurth}, \citenamefont
  {Orginos},\ and\ \citenamefont {Walker-Loud}}]{Bouchard:2016heu}%
  \BibitemOpen
  \bibfield  {author} {\bibinfo {author} {\bibfnamefont {C.}~\bibnamefont
  {Bouchard}}, \bibinfo {author} {\bibfnamefont {C.~C.}\ \bibnamefont {Chang}},
  \bibinfo {author} {\bibfnamefont {T.}~\bibnamefont {Kurth}}, \bibinfo
  {author} {\bibfnamefont {K.}~\bibnamefont {Orginos}}, \ and\ \bibinfo
  {author} {\bibfnamefont {A.}~\bibnamefont {Walker-Loud}},\ }\href@noop {} {\
  (\bibinfo {year} {2016})},\ \Eprint {http://arxiv.org/abs/1612.06963}
  {arXiv:1612.06963 [hep-lat]} \BibitemShut {NoStop}%
\bibitem [{\citenamefont {Horsley}\ \emph {et~al.}(2013)\citenamefont
  {Horsley}, \citenamefont {Najjar}, \citenamefont {Nakamura}, \citenamefont
  {Perlt}, \citenamefont {Pleiter}, \citenamefont {Rakow}, \citenamefont
  {Schierholz}, \citenamefont {Schiller}, \citenamefont {Stüben},\ and\
  \citenamefont {Zanotti}}]{Horsley:2013wqa}%
  \BibitemOpen
  \bibfield  {author} {\bibinfo {author} {\bibfnamefont {R.}~\bibnamefont
  {Horsley}}, \bibinfo {author} {\bibfnamefont {J.}~\bibnamefont {Najjar}},
  \bibinfo {author} {\bibfnamefont {Y.}~\bibnamefont {Nakamura}}, \bibinfo
  {author} {\bibfnamefont {H.}~\bibnamefont {Perlt}}, \bibinfo {author}
  {\bibfnamefont {D.}~\bibnamefont {Pleiter}}, \bibinfo {author} {\bibfnamefont
  {P.~E.~L.}\ \bibnamefont {Rakow}}, \bibinfo {author} {\bibfnamefont
  {G.}~\bibnamefont {Schierholz}}, \bibinfo {author} {\bibfnamefont
  {A.}~\bibnamefont {Schiller}}, \bibinfo {author} {\bibfnamefont
  {H.}~\bibnamefont {Stüben}}, \ and\ \bibinfo {author} {\bibfnamefont
  {J.~M.}\ \bibnamefont {Zanotti}} (\bibinfo {collaboration} {QCDSF-UKQCD}),\
  }\href@noop {} {\bibfield  {journal} {\bibinfo  {journal} {PoS}\ }\textbf
  {\bibinfo {volume} {LATTICE2013}},\ \bibinfo {pages} {249} (\bibinfo {year}
  {2013})},\ \Eprint {http://arxiv.org/abs/1311.5010} {arXiv:1311.5010
  [hep-lat]} \BibitemShut {NoStop}%
\bibitem [{\citenamefont {Bornyakov}\ \emph {et~al.}(2015)\citenamefont
  {Bornyakov} \emph {et~al.}}]{Bornyakov:2015eaa}%
  \BibitemOpen
  \bibfield  {author} {\bibinfo {author} {\bibfnamefont {V.~G.}\ \bibnamefont
  {Bornyakov}} \emph {et~al.},\ }\href@noop {} {\  (\bibinfo {year} {2015})},\
  \Eprint {http://arxiv.org/abs/1508.05916} {arXiv:1508.05916 [hep-lat]}
  \BibitemShut {NoStop}%
\bibitem [{\citenamefont {Bietenholz}\ \emph {et~al.}(2010)\citenamefont
  {Bietenholz} \emph {et~al.}}]{Bietenholz:2010jr}%
  \BibitemOpen
  \bibfield  {author} {\bibinfo {author} {\bibfnamefont {W.}~\bibnamefont
  {Bietenholz}} \emph {et~al.},\ }\href {\doibase
  10.1016/j.physletb.2010.05.067} {\bibfield  {journal} {\bibinfo  {journal}
  {Phys.~Lett.}\ }\textbf {\bibinfo {volume} {B690}},\ \bibinfo {pages} {436}
  (\bibinfo {year} {2010})},\ \Eprint {http://arxiv.org/abs/1003.1114}
  {arXiv:1003.1114 [hep-lat]} \BibitemShut {NoStop}%
\bibitem [{\citenamefont {Bietenholz}\ \emph {et~al.}(2011)\citenamefont
  {Bietenholz} \emph {et~al.}}]{Bietenholz:2011qq}%
  \BibitemOpen
  \bibfield  {author} {\bibinfo {author} {\bibfnamefont {W.}~\bibnamefont
  {Bietenholz}} \emph {et~al.},\ }\href {\doibase 10.1103/PhysRevD.84.054509}
  {\bibfield  {journal} {\bibinfo  {journal} {Phys.~Rev.}\ }\textbf {\bibinfo
  {volume} {D84}},\ \bibinfo {pages} {054509} (\bibinfo {year} {2011})},\
  \Eprint {http://arxiv.org/abs/1102.5300} {arXiv:1102.5300 [hep-lat]}
  \BibitemShut {NoStop}%
\bibitem [{\citenamefont {Green}\ \emph {et~al.}(2015)\citenamefont {Green},
  \citenamefont {Meinel}, \citenamefont {Engelhardt}, \citenamefont {Krieg},
  \citenamefont {Laeuchli}, \citenamefont {Negele}, \citenamefont {Orginos},
  \citenamefont {Pochinsky},\ and\ \citenamefont {Syritsyn}}]{Green:2015wqa}%
  \BibitemOpen
  \bibfield  {author} {\bibinfo {author} {\bibfnamefont {J.}~\bibnamefont
  {Green}}, \bibinfo {author} {\bibfnamefont {S.}~\bibnamefont {Meinel}},
  \bibinfo {author} {\bibfnamefont {M.}~\bibnamefont {Engelhardt}}, \bibinfo
  {author} {\bibfnamefont {S.}~\bibnamefont {Krieg}}, \bibinfo {author}
  {\bibfnamefont {J.}~\bibnamefont {Laeuchli}}, \bibinfo {author}
  {\bibfnamefont {J.}~\bibnamefont {Negele}}, \bibinfo {author} {\bibfnamefont
  {K.}~\bibnamefont {Orginos}}, \bibinfo {author} {\bibfnamefont
  {A.}~\bibnamefont {Pochinsky}}, \ and\ \bibinfo {author} {\bibfnamefont
  {S.}~\bibnamefont {Syritsyn}},\ }\href {\doibase 10.1103/PhysRevD.92.031501}
  {\bibfield  {journal} {\bibinfo  {journal} {Phys.~Rev.}\ }\textbf {\bibinfo
  {volume} {D92}},\ \bibinfo {pages} {031501} (\bibinfo {year} {2015})},\
  \Eprint {http://arxiv.org/abs/1505.01803} {arXiv:1505.01803 [hep-lat]}
  \BibitemShut {NoStop}%
\bibitem [{\citenamefont {Venkat}\ \emph {et~al.}(2011)\citenamefont {Venkat},
  \citenamefont {Arrington}, \citenamefont {Miller},\ and\ \citenamefont
  {Zhan}}]{Venkat:2010by}%
  \BibitemOpen
  \bibfield  {author} {\bibinfo {author} {\bibfnamefont {S.}~\bibnamefont
  {Venkat}}, \bibinfo {author} {\bibfnamefont {J.}~\bibnamefont {Arrington}},
  \bibinfo {author} {\bibfnamefont {G.~A.}\ \bibnamefont {Miller}}, \ and\
  \bibinfo {author} {\bibfnamefont {X.}~\bibnamefont {Zhan}},\ }\href {\doibase
  10.1103/PhysRevC.83.015203} {\bibfield  {journal} {\bibinfo  {journal}
  {Phys.~Rev.}\ }\textbf {\bibinfo {volume} {C83}},\ \bibinfo {pages} {015203}
  (\bibinfo {year} {2011})},\ \Eprint {http://arxiv.org/abs/1010.3629}
  {arXiv:1010.3629 [nucl-th]} \BibitemShut {NoStop}%
\bibitem [{\citenamefont {Bali}\ \emph {et~al.}(2016)\citenamefont {Bali},
  \citenamefont {Lang}, \citenamefont {Musch},\ and\ \citenamefont
  {Sch\"{a}fer}}]{Bali:2016lva}%
  \BibitemOpen
  \bibfield  {author} {\bibinfo {author} {\bibfnamefont {G.~S.}\ \bibnamefont
  {Bali}}, \bibinfo {author} {\bibfnamefont {B.}~\bibnamefont {Lang}}, \bibinfo
  {author} {\bibfnamefont {B.~U.}\ \bibnamefont {Musch}}, \ and\ \bibinfo
  {author} {\bibfnamefont {A.}~\bibnamefont {Sch\"{a}fer}},\ }\href {\doibase
  10.1103/PhysRevD.93.094515} {\bibfield  {journal} {\bibinfo  {journal}
  {Phys.~Rev.}\ }\textbf {\bibinfo {volume} {D93}},\ \bibinfo {pages} {094515}
  (\bibinfo {year} {2016})},\ \Eprint {http://arxiv.org/abs/1602.05525}
  {arXiv:1602.05525 [hep-lat]} \BibitemShut {NoStop}%
\bibitem [{\citenamefont {Nakamura}\ and\ \citenamefont
  {St{\"u}ben}(2010)}]{Nakamura:2010qh}%
  \BibitemOpen
  \bibfield  {author} {\bibinfo {author} {\bibfnamefont {Y.}~\bibnamefont
  {Nakamura}}\ and\ \bibinfo {author} {\bibfnamefont {H.}~\bibnamefont
  {St{\"u}ben}},\ }\href@noop {} {\bibfield  {journal} {\bibinfo  {journal}
  {PoS}\ }\textbf {\bibinfo {volume} {LATTICE2010}},\ \bibinfo {pages} {040}
  (\bibinfo {year} {2010})},\ \Eprint {http://arxiv.org/abs/1011.0199}
  {arXiv:1011.0199 [hep-lat]} \BibitemShut {NoStop}%
\bibitem [{\citenamefont {Boyle}(2009)}]{Boyle:2009vp}%
  \BibitemOpen
  \bibfield  {author} {\bibinfo {author} {\bibfnamefont {P.~A.}\ \bibnamefont
  {Boyle}},\ }\href {\doibase 10.1016/j.cpc.2009.08.010} {\bibfield  {journal}
  {\bibinfo  {journal} {Comput.~Phys.~Commun.}\ }\textbf {\bibinfo {volume}
  {180}},\ \bibinfo {pages} {2739} (\bibinfo {year} {2009})}\BibitemShut
  {NoStop}%
\bibitem [{\citenamefont {Edwards}\ and\ \citenamefont
  {Joo}(2005)}]{Edwards:2004sx}%
  \BibitemOpen
  \bibfield  {author} {\bibinfo {author} {\bibfnamefont {R.~G.}\ \bibnamefont
  {Edwards}}\ and\ \bibinfo {author} {\bibfnamefont {B.}~\bibnamefont {Joo}}
  (\bibinfo {collaboration} {SciDAC, LHPC, UKQCD}),\ }\href {\doibase
  10.1016/j.nuclphysbps.2004.11.254} {\bibfield  {journal} {\bibinfo  {journal}
  {Nucl.~Phys.~Proc.~Suppl.}\ }\textbf {\bibinfo {volume} {140}},\ \bibinfo
  {pages} {832} (\bibinfo {year} {2005})},\ \Eprint
  {http://arxiv.org/abs/hep-lat/0409003} {arXiv:hep-lat/0409003 [hep-lat]}
  \BibitemShut {NoStop}%
\end{thebibliography}%

\end{document}